\documentclass[twocolumn,preprintnumbers,amsmath,amssymb]{revtex4}

\usepackage{graphicx}
\usepackage{dcolumn}
\usepackage{bm}
\usepackage{psfig}

\begin{document}

\title{Negative differential Rashba effect in two-dimensional hole systems}

\author{B.~Habib, E.~Tutuc, S.~Melinte, M.~Shayegan, D.~Wasserman, S.~A.~Lyon}

\affiliation{Department of Electrical Engineering, Princeton
University, Princeton, NJ 08544, USA}

\author{R.~Winkler}
\affiliation {Institut f\"{u}r Festk\"orperphysik,
Universit\"{a}t Hannover, Appelstr.~2, D-30167
Hannover, Germany}

\date{\today}

\begin{abstract}
  We demonstrate experimentally and theoretically that
  two-dimensional (2D) heavy hole systems in single heterostructures exhibit
  a \emph{decrease} in spin-orbit interaction-induced spin splitting
  with an increase in perpendicular electric field. Using front and back
  gates, we measure the spin splitting as a function of applied
  electric field while keeping the density constant. Our results are in
  contrast to the more familiar case of 2D electrons
  where spin splitting increases with electric field.
\end{abstract}

\pacs{Valid PACS appear here}
\maketitle


In a solid that lacks inversion symmetry, the spin-orbit
interaction leads to a lifting of the spin degeneracy of the
energy bands, even in the absence of an applied magnetic field,
\emph{B}. In such a solid, the energy bands at finite wave vectors
are split into two spin subbands with different energy surfaces,
populations, and effective masses. The problem of inversion
asymmetry-induced spin splitting in two-dimensional (2D) carrier
systems in semiconductor heterojunctions and quantum wells
\cite{Stormer, Eisenstein, Wieck, Rossler} has become of renewed
interest recently \cite{Winkler03} because of their possible use
in realizing spintronic devices such as a spin field-effect
transistor \cite{Awschalom, Datta}, and for studying fundamental
phenomena such as the spin Berry phase \cite{Morpurgo, Yau}.

In 2D carrier systems confined to GaAs/AlGaAs heterostructures,
the bulk inversion asymmetry (BIA) of the zinc blende structure
and the structure inversion asymmetry (SIA) of the confining
potential contribute to the $B=0$ spin splitting \cite{Rossler,
Winkler03}. While BIA is fixed, the so called Rashba spin
splitting \cite{Rashba} due to SIA can be tuned by means of
external gates that change the perpendicular electric field
($E_\perp$) in the sample. For many years it has been assumed that
the Rashba spin splitting in 2D carrier systems is proportional to
$E_\perp$ that characterizes the inversion asymmetry of the
confining potential \cite{Rossler}. 2D holes contained in a GaAs
\emph{square} quantum well provide an example \cite{Lu}. On the
contrary, in the present work we show both experimentally and
theoretically that for heavy holes confined to a
\emph{triangular} well at the GaAs/AlGaAs interface, spin
splitting \emph{decreases} with an increase in $E_\perp$. We
demonstrate this negative differential Rashba effect by analyzing
the Shubnikov-de Haas oscillations in this system at a
constant density. We note that hole systems have recently gained
great attention for spintronics applications \cite{Pala} because
ferromagnetic (III,Mn)V compounds are intrinsically $p$ type. A
detailed understanding of the $B=0$ spin splitting in hole systems
is thus of great importance.


\begin{figure}
\centering
\includegraphics{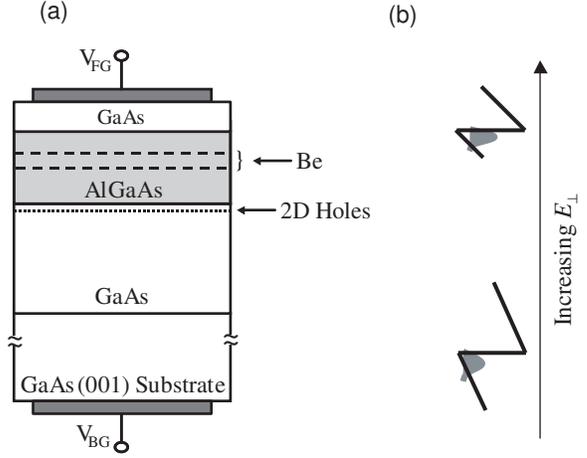}
\caption{\label{fig:sample}(a) Schematic sample cross section. (b)
Schematic demonstrating how the gate voltages change the shape of
the 2D heterostructure potential (lines) and the charge density
profile (shaded) while keeping the density constant.}
\end{figure}

The sample used in our study was grown on a GaAs (001) substrate
by molecular beam epitaxy and contains a modulation-doped 2D hole
system confined to a GaAs/AlGaAs heterostructure
[Fig.~\ref{fig:sample}(a)]. The Al$_{0.3}$Ga$_{0.7}$As/GaAs
interface is separated from a 16~nm thick Be-doped
Al$_{0.3}$Ga$_{0.7}$As layer (Be concentration of $3.5 \times
10^{18}$~cm$^{-3}$) by a 25~nm Al$_{0.3}$Ga$_{0.7}$As spacer
layer. We fabricated Hall bar samples via lithography and used
In/Zn alloyed at 440$^\circ$C for the ohmic contacts. Metal gates
were deposited on the sample's front and back to control the 2D
hole density ($p$) and $E_\perp$. The low temperature
mobility for the sample is $7.7 \times 10^4$~cm$^2$/Vs at $p = 2.3
\times 10^{11}$~cm$^{-2}$. We measured the longitudinal ($R_{xx}$)
and transverse ($R_{xy}$) magneto-resistances at $T \approx 30$~mK
via a standard low frequency lock-in technique.

In single heterostructures, where SIA is the dominant source of
spin splitting, the electric field $E_\perp$ experienced by the
carriers is determined by the density-dependent self-consistent
potential. This potential is determined, in turn, by the sample
structure (spacer layer thickness and doping, etc.), and the
applied gate biases. In our measurements, we used front and back
gate biases to change the potential's profile and hence $E_\perp$
[Fig.~\ref{fig:sample}(b)], while keeping the density constant. We
initially set the front-gate voltage ($V_\mathrm{FG}$) to $0.55$~V
and back-gate voltage ($V_\mathrm{BG}$) to $-100$~V with respect
to the 2D hole system, leading to $p = 1.84\times
10^{11}$~cm$^{-2}$, and measured $R_{xy}$ and $R_{xx}$ as a
function of $B$ between $-3$~T to $5$~T. Then at $B=1$~T, we
increased $V_\mathrm{BG}$ and noted the change in $R_{xy}$; this
change in $R_{xy}$ gives the corresponding change in density
($\Delta p$). $V_\mathrm{FG}$ was then decreased to recover the
original $R_{xy}$ and hence the original $p$. This procedure leads
to a change $\Delta E_\perp = e\Delta p/\epsilon$ in $E_\perp$
($e$ is the electron charge and $\epsilon$ is the dielectric
constant) while keeping the density constant to within 1\%.
$\Delta E_\perp$ is measured with respect to $E_\perp$ at
$V_\mathrm{FG} = 0.55$~V and $V_\mathrm{BG} = -100$~V.

\begin{figure}
\centering
\includegraphics {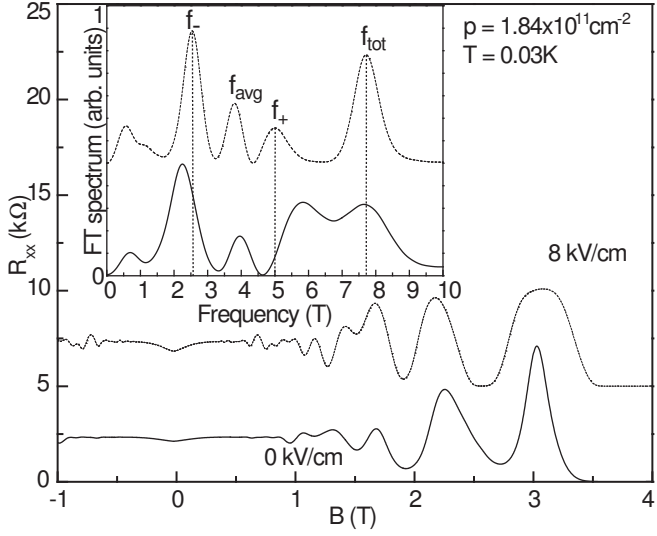}
\caption{\label{fig:fft} Observed Shubnikov-de Haas oscillations
for a 2D hole system confined to a (001) GaAs/AlGaAs single
heterostructure at two different $E_\perp$. Inset: The Fourier
spectra of these oscillations at the corresponding electric
fields. The dotted curves in the main figure and the inset are
shifted vertically for clarity.}
\end{figure}

Figure~\ref{fig:fft} shows the low-field Shubnikov-de Haas (SdH)
oscillations for two $E_\perp$ differing by 8~kV/cm. The Fourier
transform (FT) spectra of these oscillations, shown in
Fig.~\ref{fig:fft} (inset), exhibit four dominant peaks at
frequencies $f_-$, $f_\mathrm{avg}$, $f_+$, and $f_\mathrm{tot}$,
with the relation $f_\mathrm{tot} = f_+ + f_- = 2f_\mathrm{avg}$.
The $f_\mathrm{tot}$ frequency, when multiplied by $e/h$, matches
well the total 2D hole density deduced from the Hall resistance
($h$ is the Planck's constant). The two peaks at $f_-$ and $f_+$
correspond to the holes in individual spin subbands although their
positions times $e/h$ do not exactly give the spin subband
densities \cite{Winkler03, Keppeler02, Winkler00}. Nevertheless,
as discussed below, this discrepancy between $(e/h)f_{\pm}$ and
the $B = 0$ spin subband densities is minor and $\Delta f = f_+ -
f_- = f_\mathrm{tot} - 2f_-$ provides a good measure of the spin
splitting. The vertical lines in the inset of Fig.~\ref{fig:fft}
at the $f_-$ and $f_+$ peaks clearly indicate that $\Delta f$
\emph{decreases} when $\Delta E_\perp$ is increased from $0$ to
$8$~kV/cm. The vertical line at the $f_\mathrm{tot}$ peak shows
that the total hole density is held constant.

We compare the experimental data with accurate numerical
calculations of the magneto-oscillations at $B>0$. First we
perform fully self-consistent calculations of the subband
structure at $B=0$ in order to obtain the Hartree potential
$V_\mathrm{H}$ as a function of $E_\perp$ \cite{Winkler93}. We
assumed that the concentration of unintentional background
impurities in the GaAs space charge layer was $1 \times
10^{14}$~cm$^{-3}$ \cite{Stern74}. This assumption is based on our
sample parameters; we note, however, that the deduced spin
splitting is insensitive to the exact value of the background
doping. Using $V_\mathrm{H}$ we obtain the Landau fan chart for
$B>0$ from an $8 \times 8$ $\mathbf{k} \cdot \mathbf{p}$
Hamiltonian that fully takes into account the spin-orbit coupling
due to both SIA and BIA \cite{Winkler03,Trebin}. From this fan
chart we then determine the magneto-oscillations by evaluating the
density of states at the Fermi energy as a function of $B$.

\begin{figure}
\centering
\includegraphics{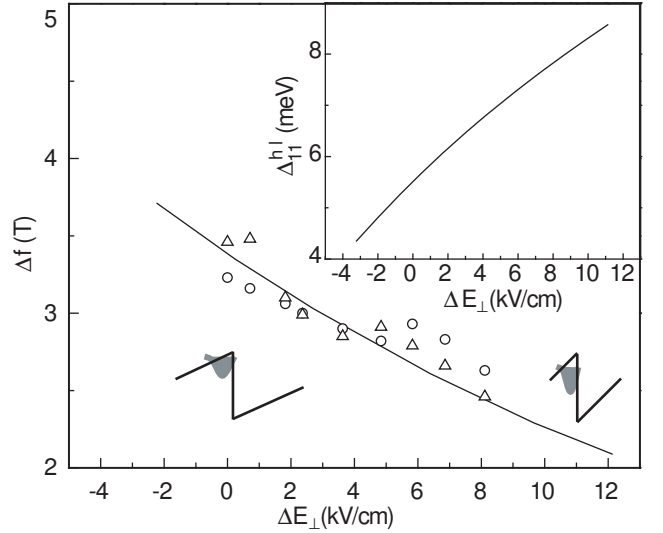}
\caption{\label{fig:ss} Spin splitting, $\Delta f$, versus the
change $\Delta E_\perp$ in the perpendicular electric field for
the measured (o = $f_\mathrm{tot} - 2f_-$, $\triangle = f_+ -
f_-$) and calculated (solid line) magneto-oscillations. The effect
of varying $E_\perp$ on the shape of the 2D heterostructure
potential (lines) and the charge density profile (shaded) is also
shown. Inset: The increase in the energy gap $\Delta_{11}^{hl}$
between the first HH and LH subbands is shown.}
\end{figure}

In Fig.~\ref{fig:ss} the experimental and calculated $\Delta f$
from the corresponding FT spectra are plotted versus $\Delta
E_\perp$. It is clear that increasing $E_\perp$ lowers the spin
splitting \cite{note}. We also verified that the calculated $B =
0$ spin splitting, defined as the difference between the spin
subband densities, shows the same negative differential Rashba
effect as $\Delta f$ obtained from the calculated
magneto-oscillations. The difference between the $B = 0$ spin
splitting times $h/e$ and $\Delta f$ is only $\lesssim 0.14$~T in
the range of $E_\perp$ shown in Fig. \ref{fig:ss}. This confirms
that the $B = 0$ spin splitting shows the same negative
differential trend.

We can understand these surprising results in the following way.
The hole states in the uppermost valence band $\Gamma_8^v$ have
the angular momentum $j=3/2$. In 2D systems, the four hole states
split into heavy-hole (HH) states with $z$ component of angular
momentum $m=\pm 3/2$ and light-hole (LH) states with $m = \pm
1/2$. Here the quantization axis is perpendicular to the 2D plane.
On the other hand, the Rashba spin-orbit coupling acts like a
$\bm{k}$-dependent effective magnetic field which is oriented in
the plane so that it favors to orient the quantization axis of the
angular momentum in-plane. However, this is not possible within
the subspace of HH states ($m=\pm 3/2$) so that --- in contrast to
$j=1/2$ electron systems
--- the Rashba spin splitting of HH states is a higher-order effect.
Neglecting anisotropic corrections, it is characterized by the
Hamiltonian \cite{Winkler03}
\begin{equation}
\label{eq:rash_HH} H^h_{\rm SO} = \beta^h_1 \, E_\perp \, i \,
(k_+^3 \sigma_- - k_-^3 \sigma_+),
\end{equation}
with $\sigma_\pm = 1/2(\sigma_x \pm i \sigma_y)$ and $k_\pm = k_x
\pm i k_y$, where $\sigma_x$ and $\sigma_y$ are the Pauli spin
matrices in the $x$ and $y$ directions respectively. Using
third-order L\"owdin perturbation theory \cite{Winkler03} we
obtain for the Rashba coefficient $\beta^h_\alpha$ of the lowest
HH subband $\alpha=1$
\begin{equation}
\label{eq:rash_HH_fak} \beta^h_1 = a\, \gamma_3
(\gamma_2+\gamma_3) \,\frac{e\hbar^4}{m_0^2} \bigg[
\frac{1}{\Delta_{11}^{hl}}
 \bigg(\frac{1}{\Delta_{12}^{hl}}
     - \frac{1}{\Delta_{12}^{hh}} \bigg)
     + \frac{1}{\Delta_{12}^{hl} \; \Delta_{12}^{hh}} \bigg],
\end{equation}
where $\gamma_2$ and $\gamma_3$ are the Luttinger parameters
\cite{Trebin} and $\Delta_{\alpha\alpha'}^{\nu\nu'} \equiv {\cal
E}_\alpha^\nu - {\cal E}_{\alpha'}^{\nu'}$ where ${\cal
E}_\alpha^h$ and ${\cal E}_\alpha^l$ are the energies of the
$\alpha$th HH and LH subband, respectively. The symbol $a$ denotes
a numerical prefactor which depends on the geometry of the
quasi-2D system. We can estimate the value of $a$ assuming an
infinitely deep rectangular quantum well which yields
$a=64/(9\pi^2)$. We see from Eq.\ (\ref{eq:rash_HH_fak}) that the
Rashba spin splitting of the HH states depends not only on the
electric field $E_\perp$ but also on the separation between the HH
and LH subbands. As can be seen in Fig.~\ref{fig:ss} (inset), the
gap between the HH-LH subbands increases with an increase in
$E_\perp$, giving rise to a decreasing Rashba coefficient
$\beta^h_1$. This result reflects the fact that a large HH-LH
splitting yields a ``rigid'' angular momentum perpendicular to the
2D plane so that the Rashba spin splitting will be suppressed.

We can estimate the effect of changing $E_\perp$ using the
well-known triangular potential approximation \cite{Stern72}. Here
we have, for the subband energies ${\cal E}_\alpha^\nu$ measured
from the valence band edge, ${\cal E}_\alpha^\nu \propto
E_\perp^{2/3}$ which implies $\beta^h_\alpha \propto
E^{-4/3}_\perp$. Therefore, we can expect from Eqs.\
(\ref{eq:rash_HH}) and (\ref{eq:rash_HH_fak}) that the Rashba spin
splitting \emph{decreases} proportional to $E_\perp^{-1/3}$ when
$E_\perp$ is increased, in agreement with our more accurate
numerical calculations.

In a previous study \cite{Winkler02} this surprising behavior of
$\beta^h_1E_\perp$ was shown as a function of density. By lowering
the density, spin splitting and $E_\perp$ both decrease but the
term $\beta^h_1E_\perp$ increases. However in the present work, by
keeping the density constant and only varying $E_\perp$, we are
able to directly demonstrate the negative differential Rashba
effect in heavy hole 2D systems confined to single GaAs/AlGaAs
heterostructures.

In conclusion, our study highlights the subtle and unexpected
dependence of the Rashba spin splitting on $E_\perp$ in 2D hole
systems. The results are important for the spintronic devices
\cite{Awschalom,Datta,Pala} whose operation relies on the tuning
of the spin splitting via applied electric field.


We thank the DOE, ARO, NSF, BMBF and the Alexander von Humboldt
Foundation for support.




\begin{thebibliography}{99}
\bibitem{note}
The slopes of the experimental and calculated curves are the
important parameters for comparison.  A reference point on the
$E_\perp$ axis, which would allow a direct comparison of the
measured and calculated results, is very hard to define because of
uncertainties in the sample parameters.
\end{thebibliography}
\end{document}